\documentclass[fleqn,12pt]{wlscirep}
 
\usepackage[utf8]{inputenc}
\usepackage{amsmath,amssymb,bbm,lmodern,graphicx}
\usepackage[section]{placeins}

\title{Effective epidemic model for COVID-19 using accumulated deaths}
\author[1,2,*]{G. Nakamura}
\author[1,2]{B. Grammaticos}
\author[1,2]{C. Deroulers}
\author[1,2]{M. Badoual}

\affil[*]{gilberto.nakamura@gmail.com}
\affil[1]{Université Paris-Saclay, CNRS/IN2P3, IJCLab, 91405 Orsay, France}
\affil[2]{Université de Paris, IJCLab, 91405 Orsay, France}

\begin{abstract}
  The severe acute respiratory syndrome COVID-19 has been in the
  center of the ongoing global health crisis in 2020. 
  The high prevalence of mild cases facilitates
  sub-notification outside hospital environments and the number of 
  those who are or have been infected remains largely unknown, leading
  to poor estimates of the crude mortality rate of the disease.
  Here we use a simple model to describe the number of accumulated
  deaths caused by COVID-19. The close connection between the proposed
  model and an approximate solution of the SIR model provides a system
  of equations whose solutions are robust estimates of epidemiological
  parameters. We find that the crude mortality varies
  between $10^{-4}$ and $10^{-3}$ depending on the severity of the
  outbreak which is lower than previous estimates obtained from
  laboratory confirmed patients. We also estimate quantities of
  practical interest such as the basic reproduction number and the
  expected number of deaths in the asymptotic limit with and without
  social distancing measures and lockdowns, which allow us to measure
  the efficiency of these interventions.  
\end{abstract}

\begin{document}
\flushbottom
\maketitle
\thispagestyle{empty}

\section{Introduction}

Outbreaks of infectious diseases have been a common occurrence
throughout history, often linked or followed by disruptions in
societies and human activities  \cite{morabia2004}. There are several
ways to measure the impact of outbreaks but death tolls are the most
relevant ones whenever the disease can threaten lives. For instance, 
32 million persons have died between 1981 and 2019 in the ongoing HIV
epidemics, 700 000 in 2018 alone \cite{whoreport2018}. Aside from this
large scale epidemic, the world has experienced several other recent
outbreaks with varying degrees of severity and scale: Zika fever,
whose symptoms are mild but can produce long-lasting effects in newborns
(microcephaly) \cite{mlakarNEJM2016,oliveiraLancet2017}; Ebola virus
disease, with a high mortality rate estimated between 20 and 75\%
\cite{ebolaNEJM2015,kerkhoveSciData2015}; Swine flu/H1N1, which became a 
pandemic in 2009-2010 although with a lower mortality rate than regular
flu \cite{simonsenPLOSMed2013}. In 2019-2020, the severe acute
respiratory syndrome COVID-19 has emerged as the most recent pandemic,
caused by the virus named SARS-CoV-2
\cite{chanLancet2020,chenLancet2020}. Due to its novelty and lack of
previous exposition, humans have no immunity against this threat,
leading to increased number of infections.  At the time of this writing, the
specifics of the pathogen transmission are still being investigated,
as well as the complete infection process once the virus enters the
host. However, it has been shown that the main human-to-human
transmission mode occurs by the spreading of contaminated droplets,
similar to other flu-like diseases \cite{chanLancet2020}. In sharp
contrast with H1N1, however, the mortality rate of COVID-19 is
estimated in the range 1-4\%, with higher ill-outcomes among persons of old
age \cite{world2020report,baudLancet2020}. This situation  creates a 
unique scenario where healthcare facilities and workers can be
overwhelmed in a short period of time, ultimately leading to
untreated patients of COVID-19 as well as other diseases
\cite{imperialcollege2020}.

To make matters even worse, asymptomatic patients can spread the
pathogen for an extended period of time, showing none or mild symptoms
during the course of the infection. As a result, laboratory tests to
detect the viral load are necessary to identify the correct number of
cases outside hospital environments. Similar to the H1N1 pandemic, the
required number of tests far  
exceeds the current amount available in most countries. Without timely
tracking of new cases, contact tracing becomes a challenging task,
hindering estimates of new cases per infection
summarized by the basic reproduction number, $\mathcal{R}_0$. The
significance 
of this parameter lies in the fact that it provides a way to glimpse
the values of the transmission rates. Those can then be used in
compartmental models -- mathematical models that describe the
evolution of epidemics assuming nearly homogeneous populations
\cite{keelingJRSoc2005,bansalJRSoc2007}. Earlier
estimates for $\mathcal{R}_0$ using epidemiological data from Wuhan, China, set
$\mathcal{R}_0$ between $1.5$ to $5.7$ without additional measures to restrict
the spreading \cite{kucharskiLancet2020,liNEJM2020,sancheCDC2020}.
With measures in place -- such as lockdown, self-isolation, and social
distancing -- $\mathcal{R}_0$ was estimated to be around $1.05$
\cite{kucharskiLancet2020}. More importantly, the insufficient number 
of tests, in addition to the long waiting time for lab results,
affects the quality of epidemic models.

In the absence of mass testings, the death toll can be used as an
alternative metric to probe the extension of the epidemic. The medical
staff can assess the cause of death from clinical reports, which may
contain test results or not, using the best of their knowledge.
Additional tests may also be appended to reports to further
specify the cause of death.

Both numbers of cases and deaths are publicly available as part of
a global effort to tackle the pandemic. {Here, we study the
  evolution of COVID-19 deaths in order to reduce the issues caused by
  the limited number of laboratory confirmed tests. 
  We show that the accumulated deaths can be effectively described by
  simple functions, namely, sigmoids whose parameters are explained in
  terms of the SIR epidemic model. The SIR model is a
  compartmental model with the following health states: susceptible,
  infective, and removed \cite{kermackProcRSocA1927}. 
  The removed state represents those who have passed away or have
  developed immunity, either by recovering from the disease or any
  other method such as vaccination.}  
The SIR model was chosen in detriment of epidemic models with
additional health states or reinfections because it is the simplest
model that addresses immunity.
{Among our results, we show that crude mortality rates can be
  computed from the parameters of sigmoids and that the rates can
  change up to one order of magnitude, depending on the 
  severity of the outbreak in a given region. }
The paper is
organized as follows. Sec. \ref{sec:data} contains the description of
the data and variables used along the text.
Sec. {\ref{sec:sir}} explains how
the SIR model is reduced from a system of differential equations to a
single non-linear differential equation,
with emphasis on the expansion around equilibrium.
Data is modeled in Sec.~\ref{sec:effective} via sigmoidal
functions, whose parameters are explained in terms of the
epidemiological parameters of the SIR model.
Time windows are addressed in Sec.~\ref{sec:timewindows}, with
special emphasis on $\mathcal{R}_0$, crude mortality rate, and
quantitative effects of the confinement.
Final comments and conclusions are listed in Sec~\ref{sec:con}.


\section{Data}
\label{sec:data}

The European Center for Disease Prevention and Control (ECDC) provides
COVID-19 data updated in a daily schedule \cite{ecdc2020}. The daily
reports portray the distribution of new confirmed cases and new deaths
presented as time series. The dataset also displays the population
size $N$ according to 2018 World Bank census for each geographical
region (see Table \ref{tab:example}).

\begin{table} 
  \caption{\label{tab:example} Example of ECDC time series for daily number
  of new cases and new deaths in France \cite{ecdc2020}.}
   \centering
    \begin{tabular}{cccccccccc}
    \hline\hline
      date  &  day &  month &  year &  cases &  deaths & country &  geoId & country code &  $N$ \\
      08/04/2020 &    8 &      4 &  2020 &   3777 &    1417 &France &    FR &                  FRA &   66987244 \\
      07/04/2020 &    7 &      4 &  2020 &   3912 &     833 &France &    FR &                  FRA &   66987244 \\
      06/04/2020 &    6 &      4 &  2020 &   1873 &     518 &France &    FR &                  FRA &   66987244 \\
      05/04/2020 &    5 &      4 &  2020 &   4267 &    1053 &France &    FR &                  FRA &   66987244 \\ 
      04/04/2020 &    4 &      4 &  2020 &   5233 &    2004 &France &    FR &                  FRA &   66987244 \\
       \hline\hline
    \end{tabular}
    
\end{table}

\begin{figure}
\centering
\includegraphics[width=0.95\textwidth]{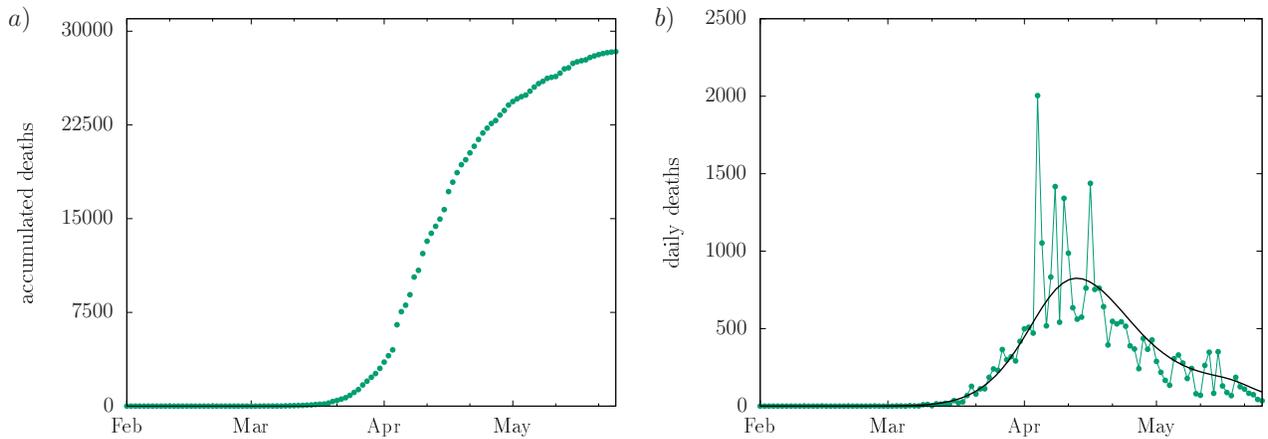}
  \caption{\label{fig1}
    Evolution of COVID-19 deaths in France.
    a) Accumulated COVID-19 deaths (circles). Countrywide measures
    to increase social distance and enforce confinement were
    introduced on March 16, and progressively removed starting from
    May 11 in a per region criteria. 
    b) Asymmetry of the daily deaths (dotted line) in France
    induced by measures to control and reduce the spreading of the
    virus among the population. (Solid line) Bézier curve for 7 days
    moving average for daily deaths.} 
\end{figure}

To better grasp the nature of the data, consider the number of
accumulated deaths in France, as shown in Fig.~\ref{fig1}. Similar
to other European countries, France was heavily affected by the
pandemic nearing the mark of 30 000 deaths, with a sharp increase
in deaths of infected patients around April-May. These deaths are
linked to infections which took place between 3 to 6 weeks
prior. On March 16, the French government implemented measures
to mitigate the propagation of the disease. The measures included
confinement of non-essential workers and temporary closures of schools
and universities, as well as commercial stores and services. The
effects of said measures did not show up immediately on the data
(see Fig.~\ref{fig1}b) but instead they appeared after some time had
passed, around 4 weeks, reducing the number of daily fatalities.

Before diving into modeling, we seek for general features that can
be used to model outbreaks. These features concern the spreading
regime of the disease among the population, excluding spatial
effects and temporal delays. As noted above, the death toll
{experienced} a rapid growth between March and April (see
Fig.~\ref{fig1}). This regime is the hallmark of epidemics and
denotes the exponential phase. In general, the growth rate in
compartmental epidemic models is summarized by $\mathcal{R}_0$ that
is the ratio of the transmission rate $\alpha$ to the removal
rate $\lambda$. As the name implies, the transmission rate dictates
the average number of persons a given infective individual typically infects
during a fixed time interval. The inverse of the removal rate is the
characteristic time in which a person remains contagious.

The inflection point is another important feature. However,
unlike small scale outbreaks, where the disease spreads
uninterruptedly among elements of a given population, the
introduction of large scale control measures forcibly
{modifies} the transmission rate. This sudden perturbation
reduces the value of the transmission rate in a short time interval,
creating an artificial inflection point.

After the exponential phase, the system relaxes toward an
equilibrium state with no infective people, with a characteristic time
scale, namely, the relaxation time $\tau$. We shall investigate the
relationship between $\tau$ with epidemiological parameters in the
next section. 
Considering these three aspects, and using Fig.~\ref{fig1} as
reference, it becomes clear that the number of accumulated deaths is a
monotonic function, with an early exponential growth only to be
replaced by a smooth relaxation towards equilibrium. Therefore, 
sigmoids appears as ideal candidates to model the data.


\section{SIR model } \label{sec:sir}

For the sake of simplicity, let us assume the spreading dynamics can
be approximated by the standard SIR model in a homogeneous population
of size $N$. The model comprises a population whose subjects can be
classified in three distinct health states, namely, susceptible,
{ infective}, and removed. The removed state { includes}  individuals that
have either died or recovered from the disease. The latter are assumed
to not be infective anymore, nor susceptible to become sick again
because of some kind of immunity. The 
fraction of individuals in each compartment is, respectively, $S(t)$,
$I(t)$, and $R(t)$, at time instant $t$. The dynamic goes as follows.
Infective subjects in the population transmit the pathogen to
susceptible ones, under adequate conditions. The transmission occurs
with rate $\alpha$, and we assume the homogeneous mixture of the
population, that is, each person in the population is statistically
equivalent to each other \cite{bansalJRSoc2007}. { Once infected, the
person remains infective for an average period $1/\lambda$, where
$\lambda$ is the removal rate}. The dynamical equations that describe
the model are:    
\begin{subequations}
  \label{eq:sir}
  \begin{align}
    \label{eq1}
    \frac{d S}{dt} &= -\alpha S(t)I(t),\\
    \label{eq2}
    \frac{d I}{dt} &= +\alpha S(t)I(t) - \lambda I(t),\\
    \label{eq3}
    \frac{dR}{dt} &= +\lambda I(t),
  \end{align}
\end{subequations}
with the constraint $S+I+R = 1$, i.e., conservation of the population
size. The model certainly simplifies or neglects recent aspects of
the COVID-19 pandemic such as differentiation between asymptomatic
and symptomatic transmission or age-dependent rates
\cite{premLancet2020,giordanoNatMed2020}.
In fact, research on the biological characteristics of the ongoing
pandemic are still ongoing \cite{gandhiNEJM2020}, but evidence indicates 
re-infections should be minimal in recovered patients
\cite{otaNatRevImmuno2020}. Therefore, we make a case for an
approximate description of the problem via the SIR model over the
inclusion of extra complexities and uncertainties, aiming to capture
dominant aspects.

The system of differential equations (\ref{eq1}-\ref{eq3}) can be
further reduced to a single first-order differential equation as
follows.  From (\ref{eq3}) and (\ref{eq1}), one finds  $ S(t) =  S_0
\textrm{e}^{ -(\alpha / \lambda )   \, R(t) } $. The constant $S_0 =
S(0) \textrm{e}^{ (\alpha / \lambda ) \, R(0) }$ depends on the
initial conditions $S(0)$ and $R(0)=1 -S(0)-I(0)$. Usually, we are
more interested in scenarios in which $R(0) =0$ and thus $S_0 = 1 -
I(0)$, similar to the onset of an emerging disease. The complete
expression for $S_0$ must be used for different initial conditions,
which can become a problem whenever the ratio $\alpha/\lambda$ is
unknown. Ignoring constant solutions, it can be shown
\cite{harkoApplMathComput2014} that the equation can be further
reduced to
\begin{equation}
  \label{eq:r}
  \frac{d R}{d t} = -\lambda S_0 \textrm{e}^{-(\alpha/\lambda ) R(t)} -
  \lambda R + \lambda,
\end{equation}
whose general solution can be obtained by quadrature
\begin{equation}
  \label{eq:t}
  t- t_0 =\frac{1}{\lambda} \int_{R(0)}^{R(t)} \frac{d r}{1 - r - S_0
    \textrm{e}^{-(\alpha/\lambda) r } }.
\end{equation}
The stationary condition in (\ref{eq:r}) gives the value $R_{\infty}$
as a solution of the transcendental equation $ R_{\infty} = 1 - S_0
\textrm{e}^{-(\alpha/\lambda)R_{\infty}}$.

\begin{figure}[!tbh]
\centering
  \includegraphics[width=0.95\textwidth]{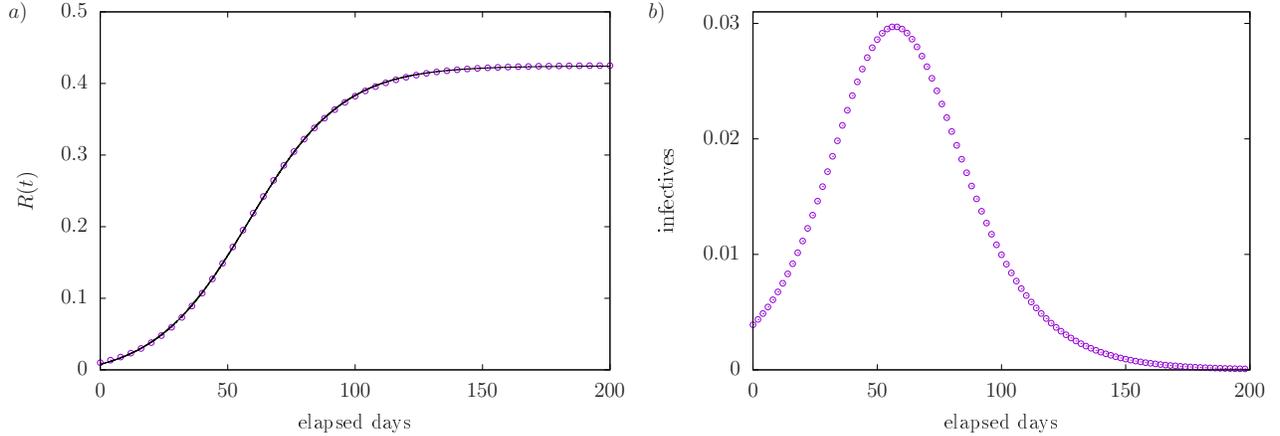}
  \caption{\label{fig:sir}
    SIR model. a) Numerical solution  of the SIR model (empty circles) with
    $\mathcal{R}_0=1.3$. A good agreement is found between the sigmoidal
    curve (\ref{eq:r_sol}) whose parameters were found by least-square
    fitting (line) and the SIR model, being less accurate for increasing
    values of $\mathcal{R}_0$. b) The curve that represents the
    infective fraction of the population in the SIR model is
    symmetrical, with peak at the center, as long as parameters
    remain constant.
  }
\end{figure}

The solution $R(t)$ can be obtained by inverting (\ref{eq:t}), for
which a general method remains unknown. To circumvent the issue, one
may expand (\ref{eq:r}) around points of interest, for example, around
$R = 0$ or $R = R_\infty$. Each expansion has advantages and
issues. The linear term in the expansion around $R = 0$ dictates the
exponential growth of $R(t)$,  with an effective rate $\alpha S_0 -
\lambda$. We use the adjective effective rather loosely here because
at some point the curve should change its curvature and converge to an
equilibrium value. Unfortunately, the competition between the early
exponential growth and relaxation towards equilibrium is often
difficult to assess near the onset of the outbreak, requiring higher
order contributions in the expansion.

Alternatively, we can get a better picture of the problem by expanding
$R(t)$ around the equilibrium. By doing so, the expansion only
requires contributions up to second order, as it already carries the
information regarding $R_{\infty}$. Furthermore, the expansion near
the equilibrium  also ensures that $\tau$ describes the dominant 
relaxation time, rather than combinations of several decay modes, each
one with its own timescale. Define $\delta R(t) =
R_{\infty} - R(t) \geqslant 0$. The expansion of (\ref{eq:r}) near
$\delta R \ll 1$, 
together with the transcendental equation for $R_\infty$, gives
\begin{equation}
  \label{eq:approx}
  \frac{1}{\lambda}\frac{d }{d t}\delta R = - \left[1 -
    (1-R_{\infty})\frac{\alpha}{\lambda}\right] \delta R
  + \frac{1-R_{\infty}}{2}\,\left(\frac{\alpha}{\lambda}\right)^2
  \delta R^2 + o(\delta R^3).
\end{equation}
Keeping terms up to $o(\delta R^2)$, one converts (\ref{eq:approx})
into a Bernoulli equation 
with relaxation time $\tau = [\lambda - \alpha (1-
R_{\infty})]^{-1}$. Solving for $\delta R$ and transforming back to $R$,
we find the approximate solution (see Fig.~{\ref{fig:sir}})
\begin{equation}
  \label{eq:r_sol}
  R(t) = \frac{R_{\infty}-A \,
  \textrm{e}^{-t/\tau}}{\;\;\;\; 1+B\, \textrm{e}^{-t/\tau}} \;, 
\end{equation}
with $A = R_{\infty} -  (1+B)R(0) $, $B = c_0 \tau / z_0 $, $\, z_0 =
[R_{\infty}-R(0)]^{-1}- c_0 \tau$, and $c_0 =
(\lambda/2)(1-R_{\infty})(\alpha/\lambda)^2$.

\section{Effective model}
\label{sec:effective}

  The sigmoidal expression in (\ref{eq:r_sol}) satisfies the requirements 
  listed in Sec.~\ref{sec:data} and it is an excellent candidate to
  model the data of accumulated deaths. However, the removed
  compartment corresponding to $R(t)$ holds both recovered and deceased
  fractions of the population, i.e., all the infected who are unable
  to spread the disease. We can simplify this issue by assuming the
  existence of a simple relation between $R(t)$ and the accumulated
  number of deaths divided by the population size, $g(t)$, at the time
  instant $t$. To keep the model as simple as possible, we neglect
  temporal delays and impose
  \begin{equation}
    g(t) = f R(t),
  \end{equation}
  where the crude mortality rate $f$ is the ratio between deceased and
  infected. As such, it also can be used as an estimator for the
  likelihood to die after contracting the disease.

  The equilibrium value $g_{\infty}$ varies
  from country to country and can be used to characterize the
  outbreak, more specifically, to assess the impact of the outbreak in
  the afflicted population. The monotonic nature of cumulative
  quantities together with the upper bound  $g_{\infty}$ restrict the
  possible functional forms for $g(t)$. Sigmoids are
  natural candidates to describe $g(t)$  since they are bounded and
  monotonic. Here we  consider the following
  expression in tandem with 
  (\ref{eq:r_sol}):
  \begin{equation}
    \label{eq:eff}
    g_{\textrm{eff}}(t) \equiv \frac{g_{\infty} - a\, \textrm{e}^{-t/\tau}
    }{1+b\, \textrm{e}^{-t/\tau}}.
  \end{equation}
  The timescale $\tau$ dictates the relaxation of $g(t)$ towards
  $g_{\infty}$, {with inflection at $t_c=\tau \ln b$}.
  Equation~(\ref{eq:eff}) has 3 or 4 parameters depending on whether
  $g_{\textrm{eff}}(t)$ must pass by the initial entry $g_{\textrm{eff}}(0)=g(0)=g_0$ or not.
  The former implies the constraint $a = g_{\infty} - g_0(1+b)$,
  whereas the later $g_{\textrm{eff}}(0) = (g_{\infty}-a)/(1+b)$
  with $a \leqslant g_{\infty}$. Since the data are noisy,
  we do not require the fitting curve to pass by $g_0$.

  
  The relationship between $g(t)$ and $R(t)$ is consistent if the
  following equations hold:
\begin{subequations}
  \begin{align}
    \label{eq:sys1}
    \frac{1}{\tau} & = \lambda - \alpha \left( 1-\frac{g_\infty}{f}\right),\\
    \label{eq:sys2}
    g_\infty &= f -f S_0 \textrm{e}^{-(\alpha/\lambda f) g_\infty},\\
    \label{eq:sys3}
    b& =\frac{(f-g_\infty)[g_\infty -g_{\textrm{eff}}(0)]\alpha^2}{2\omega\lambda^2 f^2-
      (f-g_\infty)[g_\infty -g_{\textrm{eff}}(0)]\alpha^2} 
  \end{align}
\end{subequations}
The system of equations~(\ref{eq:sys1}-\ref{eq:sys3}) connects the
epidemiological parameters $(\alpha, \lambda, f, S_0)$ with the
parameters $(\tau, g_{\infty}, b)$ of the sigmoid
curve~(\ref{eq:r_sol}), which can be estimated by a least-square
fitting procedure. However, there are more
variables than equations so at least one epidemiological variable
must be {fixed}. There are two reasonable
choices, namely, either $S_0$ or $\lambda$. The first choice should be
selected if the data includes the onset of the outbreak because $S_0
\approx 1$, since the majority of the population should be in the
susceptible state at the early stage of the epidemic. However, if the
beginning of the outbreak is unknown, the assumption $S_0 \approx 1$ is no longer valid.
Alternatively, one may consider an estimate for the removal
rate $\lambda$, or its probability distribution. In this case, the
input parameter to solve (\ref{eq:sys1}-\ref{eq:sys3}) depends solely
on the characteristics of the disease and local demographics. It does
not require details concerning the disease spreading nor the moment in
which the outbreak begins. Therefore, evidence-based values for
$\lambda$ from patient data are far more suitable for the purposes of this
study, and shall be used hereafter.


\section{Time windows}
\label{sec:timewindows}
  
\begin{figure}
\centering
  \includegraphics[width=0.9\columnwidth]{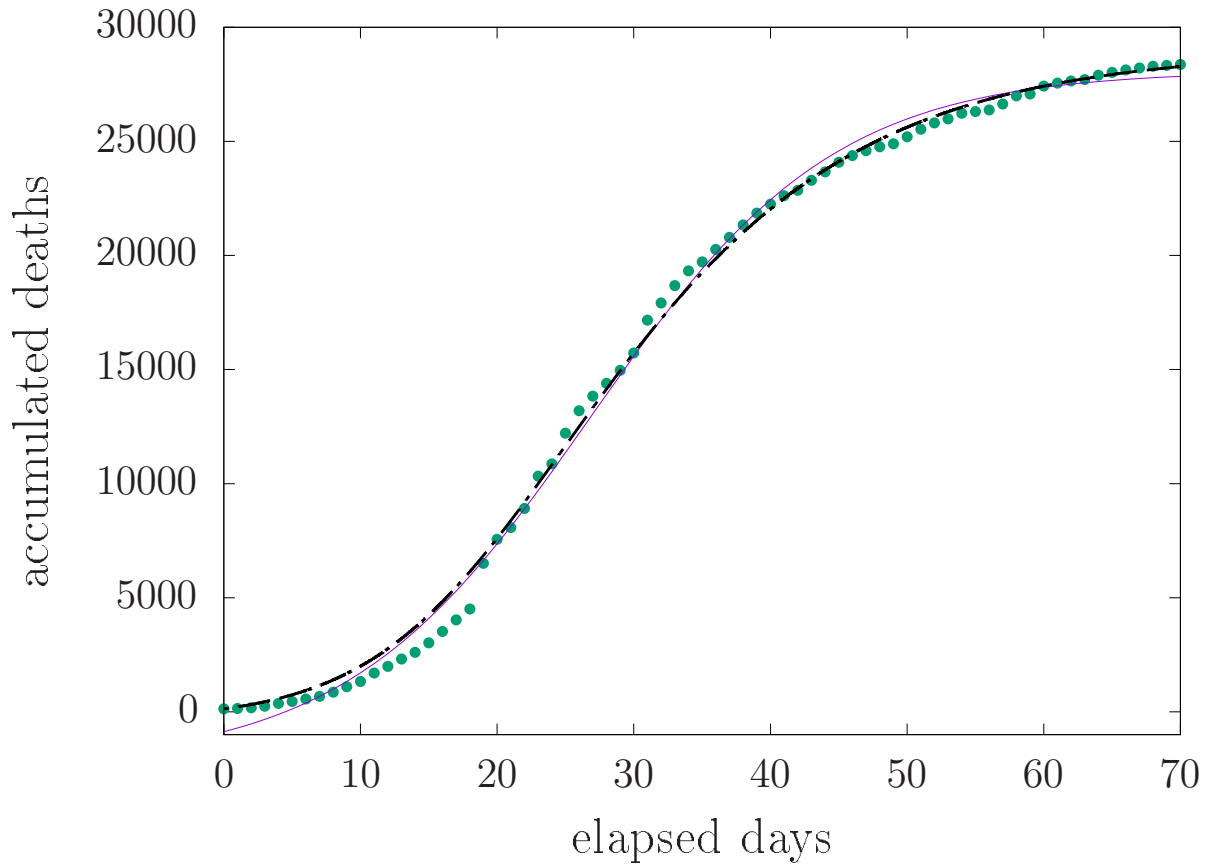}
  \caption{\label{fig:singlewindow} Issues when one tries to model the
    data with a single time window
    using French COVID-19 deaths from March 16 to May 25. 
    Parameter optimization of the SIR model (dashed line) agrees
    well with input data (circles), especially after April 5. The
    optimization neglects the effects of control measures to reduce
    the transmission rate, while reducing the removal
    rate to $\lambda = 0.103$ day$^{-1}$. The fit by a sigmoidal curve (solid
    line) underestimates $g_{\infty}$ and the curve turns negative
    in early March.      }
\end{figure}

Fig.~\ref{fig:singlewindow} shows the death toll in France from
March 16 to May 25. Unlike the theoretical SIR model, agreement
between data and the sigmoidal fit (performed with a standard
non-linear least-square fitting procedure) is poor. The fitted curve becomes
negative in early March and converges to a different equilibrium
value. Thus, we must refute the sigmoid to describe the death toll
in the complete time interval. Alternatively, the parameter
optimization of the SIR model agrees well with the input
data, except for the first 20 days. However, the optimization
returns $\lambda_{\textrm{opt}}= 0.103$ days$^{-1}$ which does not match the
signature value for COVID-19 $\lambda = 0.2$ day$^{-1}$.
If the value $\lambda=0.2$ days$^{-1}$ is held fixed during the
optimization (not shown), then the optimal set of parameters become
highly sensitive to the initial guess of the optimization algorithm.

The observation above inquires whether the SIR model describes the data
or not. However, this issue can be understood by inspecting the
number of daily deaths (see Fig.~{\ref{fig1}}a). An asymmetry is
observed around the peak of daily deaths (see
Fig.~{\ref{fig1}}b), which is in sharp contrast with the
corresponding curve in the theoretical SIR model (see
Fig.~{\ref{fig:sir}}b). The asymmetry can be explained by a 
variation in the transmission rate. Unlike other recent epidemics,
several countries have adopted control measures such as lockdowns of
non-essential workers, flight and other travel restrictions. The
efforts effectively reduce the transmission rate of COVID-19 once
they are in place. Thus, the data must be divided in non-overlapping
time windows, each with its own set of epidemiological parameters.

\begin{figure}
\centering
  \includegraphics[width=0.9\columnwidth]{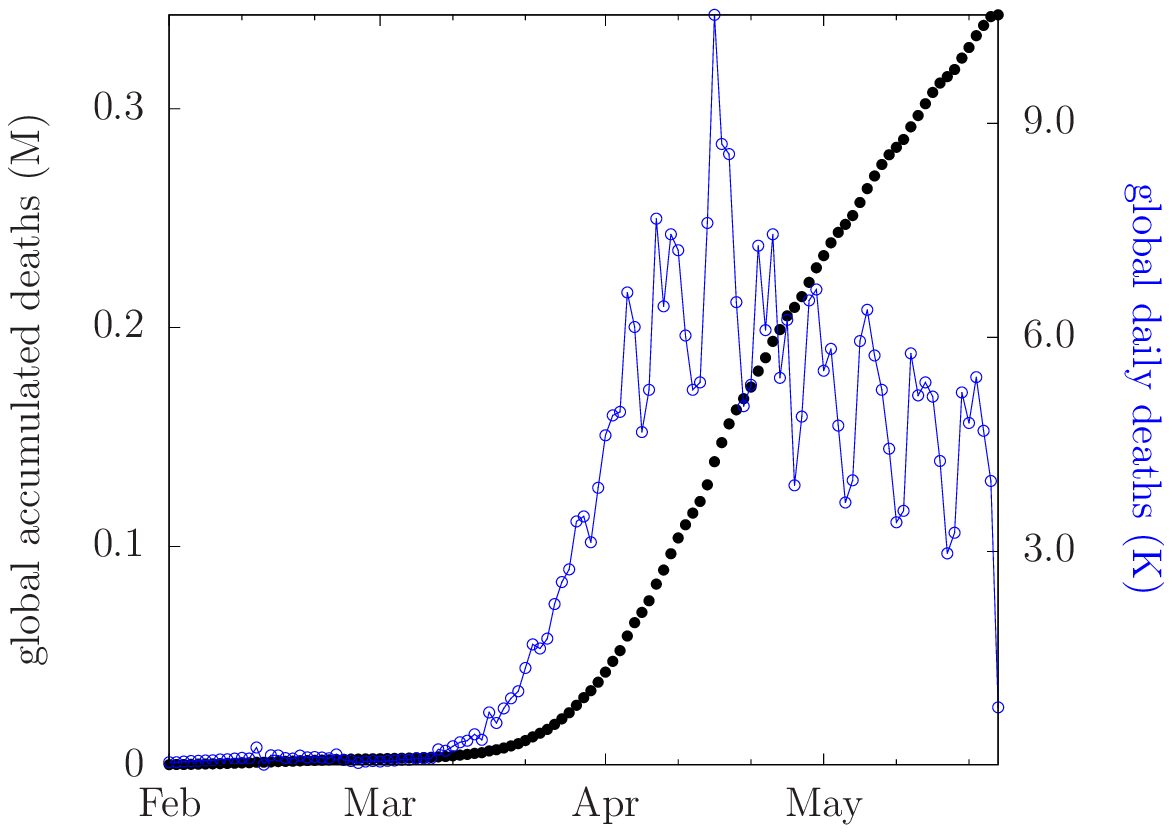}
  \caption{\label{fig:world} Global evolution of COVID-19.
    Accumulated deaths in millions (full circles).  Daily deaths in
    the thousands (line with circles). }
\end{figure}

\begin{figure}
\centering
  \includegraphics[width=0.9\columnwidth]{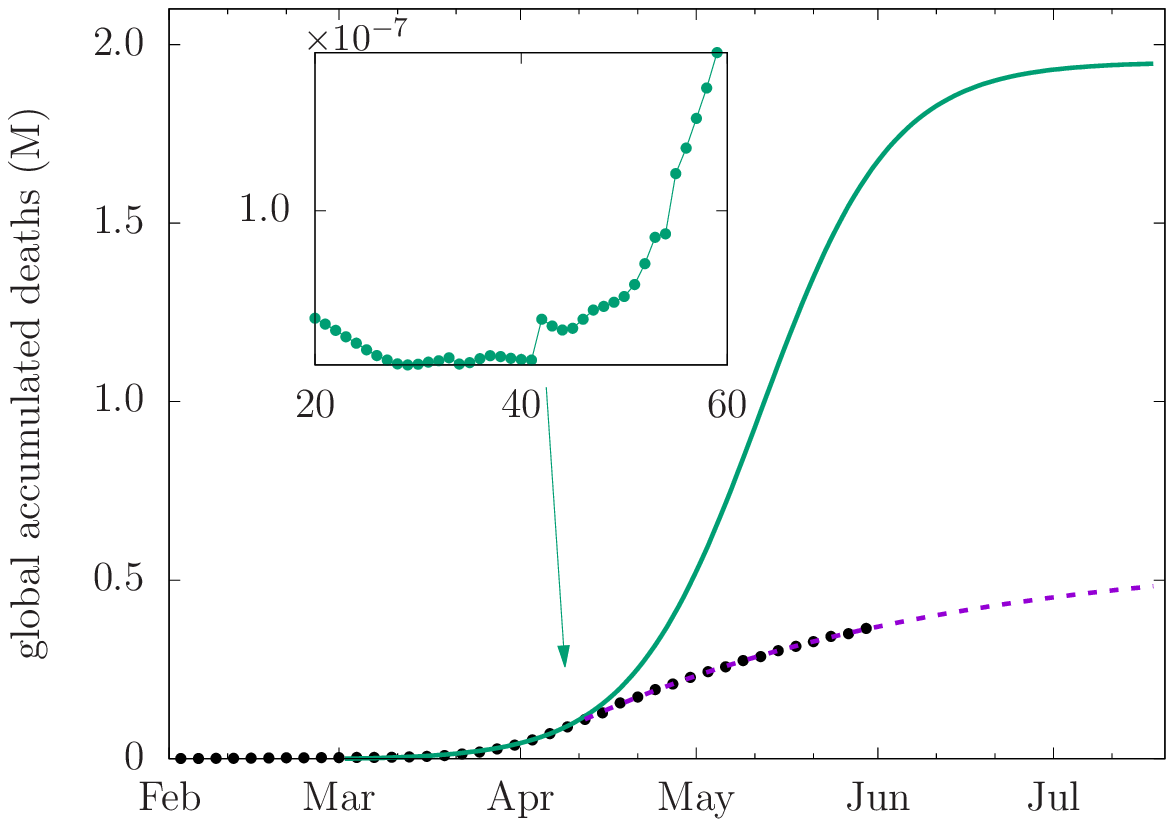}
  \caption{\label{fig:world2}
    {
      Time windows and the global evolution
      of COVID-19. The data (circles) are separated in two consecutive
      time windows (February 29 - April 9) and (April 10 - May 31). 
      The fit (\ref{eq:eff}) in the first time window (solid line)
      predicts over $2.0$ million deaths in a scenario without control
      measures. With control measures in place in the second time window,
      the death toll converges to $0.6$ million deaths (dashed line).
      (inset) residue per degree of freedom $r^2$.  The size of the
      first time window is chosen as to maximize the adherence of
      the fitting curve (minimize $r^2$) while increasing the number
      of data points. $r^2$ forms a plateau for 
      sizes between $29$ to $41$ days followed by a sudden increase
      for $42$ days, indicating that $41$ days is the optimal size of
      the time window (February 29 - April 9). }  
  }
\end{figure}

As a first example of data where the necessity to consider several
time windows is obvious, we consider the global deaths
between February 29 and May 31 (see Fig.~\ref{fig:world}). This case
is interesting because countries afflicted by the epidemic have
implemented various control measures at different schedules, of
varying effectiveness. Thus, reduction of the global transmission
rate cannot be pinpointed to a single day or week a priori. Similar
to  Fig.~\ref{fig1}, the number of daily deaths exhibits an asymmetry, 
with center in mid-April, indicating the approximate interval in
which the global transmission rate changes. European countries with
most cases at the time (France, Italy, Spain)  introduced lockdown
and other strategies in mid-March, with others following
shortly. Afterwards, the daily number of deaths starts to reduce,
followed by an oscillatory pattern (7 days period) likely tied to
work routines of medical staff and death cases reports. 

In the following, we decide to divide the observation time in two
consecutive time windows. Let us explain how we determine the optimal
separation time (optimal duration of the first time window).
Let $g_k$ be the fraction of the global population that dies due to
complications caused by COVID-19, starting from February 29. The
index $k=0,1,\ldots, m-1$ indicates the number of elapsed days in the
time window with duration $m$. The data are fitted via
(\ref{eq:eff}) with parameters $g_{\infty}, a, b $ and $\tau$, using
trust region reflective algorithm (Python/Scipy). 
It is convenient to fit the data using  $\tau^{-1}$ instead of
$\tau$ and a finite interval $0 \leqslant \tau^{-1} \leqslant  1$,
so the infection lasts at least one day in the mathematical model. In
the fitting procedure, we use bounds for $b$ which 
depends on the time interval $t_c$ between the inflection point
and the initial entry, more specifically, $t_c   = \tau \ln
b$. For $t_c > 0$ we restrict the search to $t_c/\tau  < 10 $ so
that  $b \sim o(10^{4})$; for $t_c < 0$, i.e., starting the
counting after the inflection point, the parameter space of $b$ is
limited between 0 and 1. Thus, we set the $0 < b \leqslant
10^{4}$. The remaining parameters are restricted to $0 \leqslant
g_{\infty}, a \leqslant 1$.

The quality of the fit is quantified by the square residue divided by
the size of time window, namely, $r^2 = (1/m)\sum_{k=0}^{m-1}[g_k -
  g_{\textrm{eff}}(k)]^2$. The first time window comprises $m$ 
consecutive days that minimize $r^2$ while increasing $m$, as
indicated by the inset in Fig.~\ref{fig:world2}, from February 29 to
April 10, whereas days starting from the $m+1$ day are put into the
second time window. The curve fit in the first interval corresponds to
a scenario in which 
  control measures were not implemented outside China and South
  Korea. These two time windows are consistent with the asymmetry observed
  in the number of daily deaths in Fig.~\ref{fig:world}, whose peak
  lies in mid-April. The fit converges to near $2.0$ million deaths,
  significantly above the equilibrium value ($0.6$ million deaths) in
  the second time window, from April 11 to May 31. We stress that the
  number of fatalities avoided ($1.4$ million) highlights the
  effectiveness of lockdowns and other control measures.

  The curve fitting in the first time window, which includes the phase
  with exponential growth, can be challenging though. Even more so if
  the fitting data contains a large number of entries around the
  inflection point induced by the introduction of control
  measures. This inflection point is not the natural inflection point
  which occurs in an uncontrolled epidemic with constant parameters. Due to the significance of
  inflection points in the curve fitting, the induced inflection
  point can be responsible for the artificially low value of the
  equilibrium value $g_{\infty}$ in the first time window. Also, early
  data often contain far less entries, thus being more susceptible to fluctuations. 

\begin{figure}
  \centering
    \includegraphics[width=0.9\columnwidth]{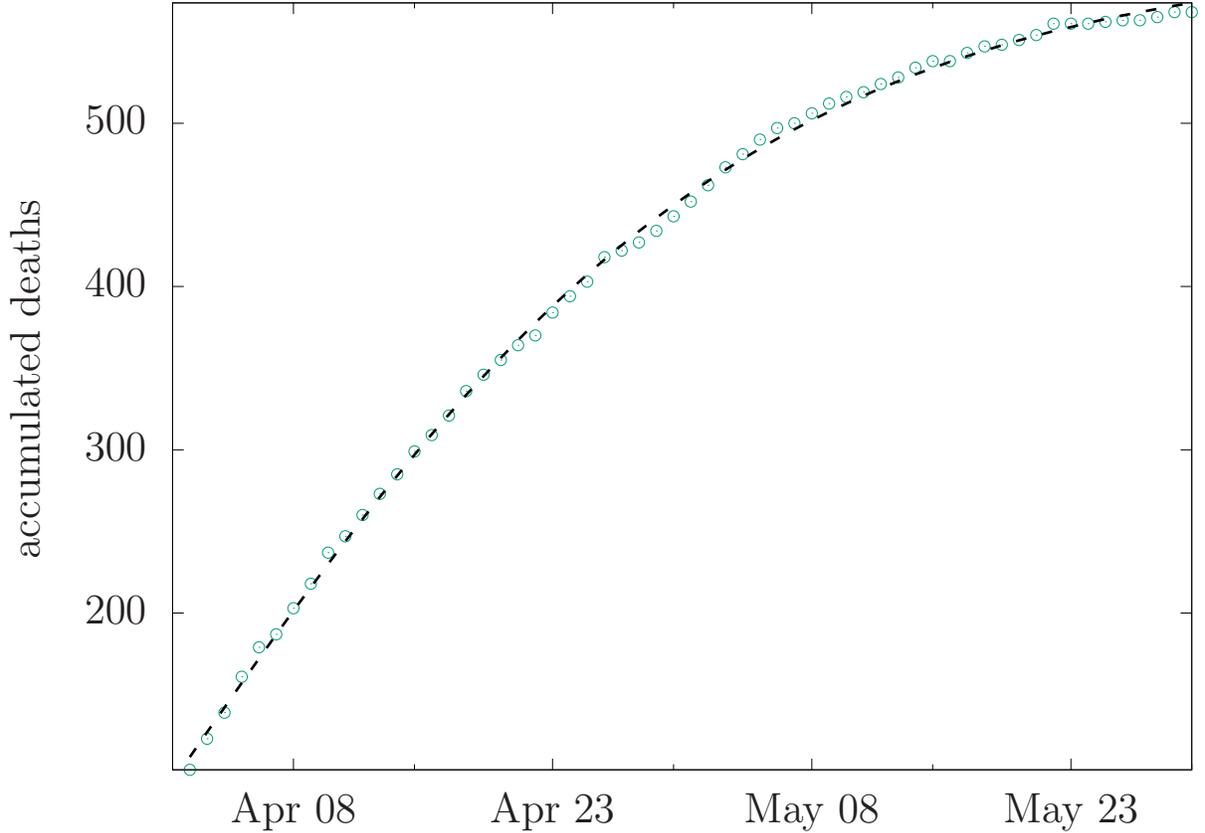}
    \caption{\label{fig:denmark} COVID-19 deaths in Denmark from April
      4 to May 31. The sigmoidal fitting (line) of input data (cicles)
      followed by the resolution of (\ref{eq:sys1}-\ref{eq:sys2})
      produce $\mathcal{R}_0 = 0.81$ and $f=6.86\times 10^{-4}$. 
    }
  \end{figure}

  Conversely, the presence of inflection points caused by the
  introduction of control measures greatly simplifies the fitting
  procedure in the second time window. For instance,
  Fig.~\ref{fig:denmark} depicts the death toll in Denmark after April
  3, within the second time window.
  The fitting (\ref{eq:eff}) is in excellent agreement with data for
  $g_{\infty} = (1.03 \pm 0.01) \times 10^{-4}$ (approximately $600$
  deaths) and $\tau = 15.84 \pm 0.65$ days.
  Next, we solve the system
  (\ref{eq:sys1}-\ref{eq:sys3}) for $\lambda$ distributed according to
  some probability distribution, centered around $\lambda = 0.2$
  days$^{-1}$. In practical terms, the width of the probability
  distribution forms the ground to compute uncertainties of the
  transmission and crude mortality rates.  However, the finer details
  of the distribution of $\lambda$ remain unknown so we resort to {a}
  uniform distribution {for virus shedding}, whose interval lies
  between 3 and 7 days \cite{wolfelNature2020,premLancet2020}. By doing
  so, we overestimate the uncertainties of the remaining
  epidemiological parameters of the SIR model
  {and find $\mathcal{R}_0 = 0.81 \pm 0.07$.
    Our estimate for the crude mortality rate $f = (0.68 \pm 0.15)
  \times 10^{-4}$ is compatible with the estimate $f_{\textrm{DNK}} = 8.2 \times
  10^{-4} $ ( confidence interval: $[5.9  - 15.4] \times  10^{-4} $) obtained by
  screening antibodies of $20 640$ blood donors below the age of
  70, in Denmark \cite{erikstrupMedrxiv2020}. 
  Also, the asymptotic value $R_{\infty} = g_{\infty}/f = 0.15
  \pm 0.03$ places the fraction of infected well below the threshold
  ($60\%$) for herd immunity.  Thus, a new wave of infections is
  likely to occur if the disease becomes seasonal unless a vaccine
  becomes available or social distancing measures remain in place. }

  \begin{figure}
  \centering
    \includegraphics[width=0.9\columnwidth]{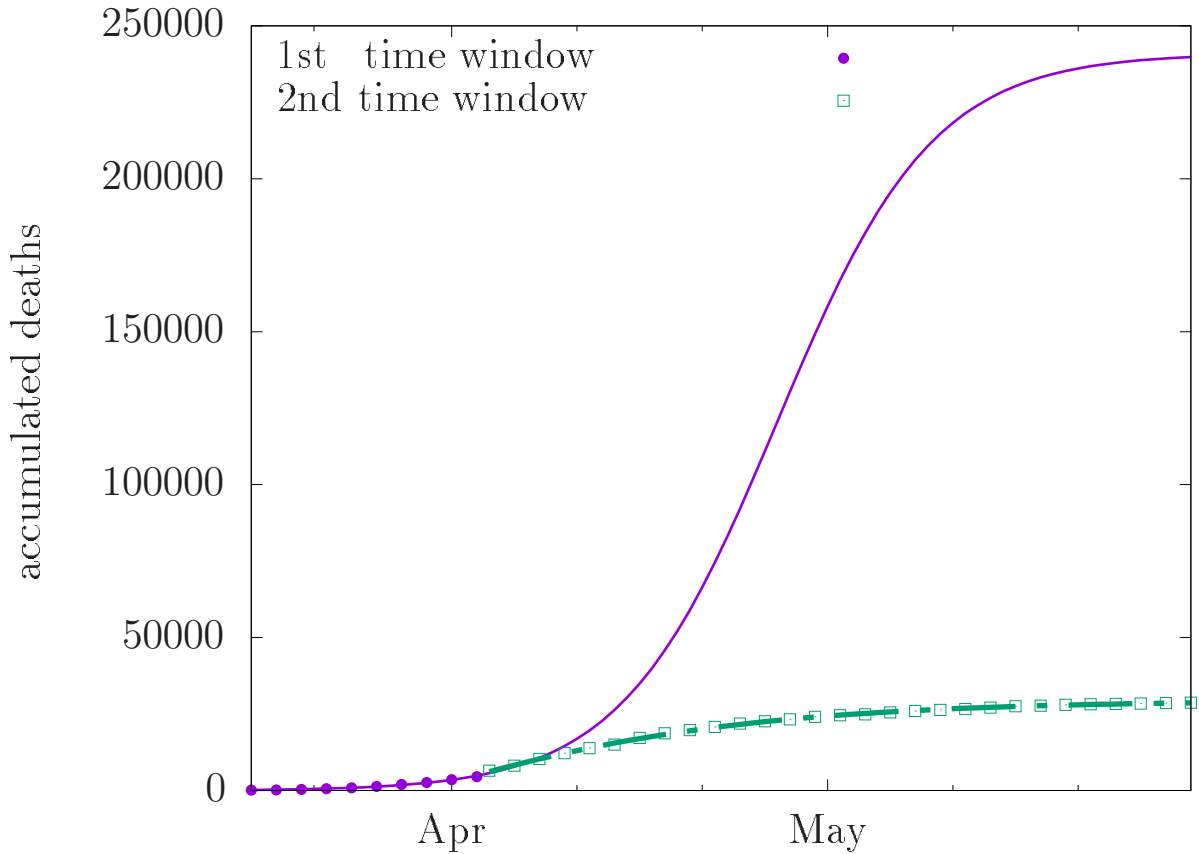}
    \caption{\label{fig:france} Evolution of COVID-19 in France
      between March 16 and May 31. Reduced number of data points for
      clarity. Data (full circles) and sigmoidal fit (solid line)
      between March 16 and April 3, with $\mathcal{R}_{0,1} = 3.25 $ and
      crude mortality rate $f_1=4.23 \times 10^{-3}$.  The effects of
      confinement emerge shortly after April 3, with a significant
      reduction in $\mathcal{R}_{0,2}=0.74$, as indicated by the
      sigmoidal fit (dashed line) of data in the second time window
      (empty squares). The mortality rate remains nearly unchanged
      $f_2=4.60 \times 10^{-3}$.
    }
  \end{figure}

  We can now move to more complicated cases, in which the data themselves
  contains artifacts and the fitting procedure can be tricky. That is
  the case of France, for instance. The lockdown was issued on March
  16, and unaccounted deaths in nursing homes were added to the
  official statistics on April 3 and 4, producing a large fluctuation
  in the number of daily deaths. In this case, we 
  separate the time windows according to the number of deaths. The
  first time window lies between 100 and 5000 deaths (March 16 - April
  3), whereas the remaining days until May 31 comprises the second
  time window, as shown in Fig.~\ref{fig:france}. 

  The curve  fitting in the second time window is far more stable and
  insensitive to initial guesses, or bounds, with equilibrium death
  toll nearing  $29440 \pm 158$ and time scale $\tau = 15.12\pm
  0.72$ days, which approaches the typical recovery time for mild
  COVID-19 infections. The solution of the system
  (\ref{eq:sys1}-\ref{eq:sys3}) returns $\mathcal{R}_{0,2} = 0.74 \pm
  0.08 < 1$ in agreement with the decline of new infections, where
  the notation $\mathcal{R}_{0,k}$ with $k=1$ and $2$ refers to the
  first and second time windows, respectively. In addition, the
  crude mortality rate in the second time window reads $f_2 =
  (4.67\pm 1.16)\times 10^{-3}$, one order of magnitude higher than
  Denmark or Germany (see Table.~\ref{tab:params}).  

  \begin{table}
    \caption{\label{tab:params} Parameters of the SIR model and
      sigmoid from April 4 to May 31 for selected
      countries, {using a uniform distribution for $\lambda$.}  }
\centering
      \begin{tabular}{lccccc}
      \hline\hline
        & $\alpha$ & $\mathcal{R}_0$  & $ f \, (10^{-3}) $& $ S_0$ & $
        g_{\infty}    \, (10^{-4}) $\\  
        {France} & $0.17 \pm 0.07$ & $0.74 \pm 0.08$ & $4.67
        \pm 1.16$ & $0.97 \pm 0.02$ & $4.40 \pm 0.02$  \\ 
        {Italy} & $0.19 \pm 0.07$ & $0.84 \pm 0.07$ & $ 4.17
        \pm 0.78$ & $0.96 \pm 0.02$ & $5.75 \pm 0.01$ \\ 
        {Spain} & $0.19 \pm 0.07$ & $0.83 \pm 0.07$ & $2.31 \pm
        0.63$ & $0.90 \pm 0.06$ & $6.08 \pm 0.04$ \\ 
        {UK} & $0.18 \pm 0.07$ & $0.82 \pm 0.07$
        & $5.86 \pm 1.08$ & $ 0.97 \pm 0.01$ & $6.26 \pm
        0.05$ \\  
        {Germany} & $0.18 \pm 0.07$ & $0.83 \pm 0.07$ & $0.39
        \pm 0.11$ & $0.90 \pm 0.06$ & $1.05 \pm 0.01$ \\ 
        {Sweden} & $0.19 \pm 0.07$ & $0.85 \pm 0.06$ & $4.84 \pm
        0.77$ & $0.98 \pm 0.01$ & $5.08 \pm 0.18$ \\ 
        {Denmark} & $0.18 \pm 0.07$ & $0.81 \pm 0.07$ & $0.68
        \pm 0.15$ & $0.95 \pm 0.02$ & $1.03 \pm 0.01$ \\ 
        {Belgium}\footnote{From April 11 to May 31} & $0.17 \pm 0.07$
        & $0.73 \pm 0.08$ & $9.97 \pm 2.55$ & $0.97 \pm 0.02$ & $8.54
        \pm 0.05$   \\
        \hline\hline
      \end{tabular}
  \end{table}

  The analysis in the first time window requires more care. The
  fitting procedure requires bounds, otherwise the fit may
  incorrectly converge to an equilibrium value that is much lower than
  the one obtained for the second time window. In addition, multiple 
  solutions can be found for (\ref{eq:sys1}-\ref{eq:sys3}), several of
  which are not realistic. Instead of using brute force, it is far
  more convenient to approximate the solution and set $S_0 = 0.99$.
  In such case, (\ref{eq:sys3}) is discarded  and the remaining
  equations produce $\mathcal{R}_{0,1} = 3.25 \pm 1.71 $. The crude mortality rate in the 
  first time window, $f_1 = (4.23 \pm 0.58)\times 10^{-3} $, shares
  the same order of magnitude as $f_2$, indicating that overall the
  French {health} care system remained  responsive throughout the
  epidemic. If control measures had not been implemented, the expected
  number of deaths would have soared and reached 240~000, with equilibrium
  infected fraction of population $R_{\infty}=0.96$, assuming the
  mortality rate would have remained roughly the same.


\section{Conclusion}
\label{sec:con}


The tragic developments in the COVID-19 pandemic have exposed 
flawed aspects in protocols used to assess large scale
epidemics. Despite the various improvements in the global capacity to
produce laboratory tests, usually based on reverse-transcription
polymerase chain reaction (rRT-PCR), the majority of afflicted countries were
unable to {enforce} mass testing policies. This shortcoming has also been
experienced in 2002 with the SARS epidemic but in a smaller scale. The
lack of  
mass testing contributed in keeping the number of cases unknown,
affecting the accuracy of disease spreading models. In this paper, we
resort to the death toll from April 4 to May 31 instead of reported
cases as our primary data 
source as death certificates are mandatory and may
contain other medical assessments linking the cause of death with
COVID-19 outbreaks. We emphasize this methodology is not immune itself
from sub-notification, nor unforeseen delays on death certificates.

We model the death toll via the sigmoid curve in (\ref{eq:eff}),
which requires the parameters $g_{\infty}$ (capacity) and $\tau$
(time scale). The capacity limits the growth of the outbreak, which is
expected to vary with geographic region, healthcare quality, and
efforts to control the spreading of the virus.
Together with the crude mortality
rate $f$, which also depends on healthcare and population
demographics, they give access to a far more credible
estimate for the infected fraction of the population $R_{\infty} =
g_\infty / f$. To keep the model as simple as possible, we also
neglect temporal delays between $g(t)$ and $R(t)$. The inclusion of
said delay may introduce additional effects, but those are not
expected to be dominant in this case since spatial effects are not
being investigated.

The advantage of our approach relies on the curve fitting of a
monotonic curve (death toll) to a sigmoid in sharp contrast to complex
optimization schemes for models with multiple health states 
\cite{premLancet2020,giordanoNatMed2020}. Epidemiological parameters
are extracted by solving the algebraic system of equations
(\ref{eq:sys1}-\ref{eq:sys3}) for transmission rate, crude mortality
rate and initial condition, respectively, $\alpha, f,$ and $S_0$.  
Alternatively, the epidemiological parameters calculated from the curve
fitting can be used as educated guesses  in optimization 
algorithms reducing the likelihood of obtaining unrealistic optimal
solutions. We stress that if $f$ is known, then the fraction of
reported cases can be easily computed $R(t) = g(t)/f$. We find that
$f$ ranges from $o(10^{-4})$ to $o(10^{-3})$, depending on geographic
location. Such values lie well below recent the estimate
\cite{onderJAMA2020} using only rRT-PCR tests  from hospitalized 
cases ($7.2\%$), whereas it is more in line with the values obtained from
antibody screening with large sample size in Denmark
\cite{erikstrupMedrxiv2020}. Interestingly, countries lightly
afflicted by the epidemic exhibit lower $f$ {even though the
  transmission of the virus is similar to neighbouring countries.}
Taking Germany or Denmark as examples, we see that both have
$f \sim o(10^{-4})$ while maintaining $\mathcal{R}_0$ compatible with
other European countries. { The result is likely attributed to
  healthcare resources and services either being more accessible or
  efficient, including early surveillance.}

{
The sigmoidal fitting becomes far more reliable once the outbreak has
been active for some time as the data starts to move away from the
exponential phase. The distance from the inflection point is another
important factor that affects the quality of the fit, especially the
value of $g_{\infty}$. The introduction of lockdowns and other social
distancing measures reduce the transmission rate and effectively
create a new inflection point, which is different from the 
one expected without control measures. Thus, the fitting curve may
converge to an incorrect equilibrium value if the fitting data
include points around the induced inflection point. This can only
be solved by introducing disjoint time windows for different regimes
of $\alpha$ as Fig.~\ref{fig:france} shows.
The difference between the equilibrium values of $g_{\infty}$ from
each time window returns the fraction of avoided deaths and it
evaluates the effectiveness of control measures and policies. For
instance, approximately $210\, 000$ deaths or about seven times the
current projection have been avoided in France. This is compatible
with the large decrease in the basic reproduction number, placing it
below the endemic threshold. However, some care is needed in
interpreting these estimates as our analysis considers only two time
windows and therefore  does not anticipates second waves of infections.  
}

Concerning deviations in parameters estimated through
(\ref{eq:sys1}-\ref{eq:sys3}), they can be tracked to the
uncertainties in the removal rate. In general, the inverse of the
removal rate describes the average time required for an 
infective person to change to the removed state.
For COVID-19, virus shedding occurs more prominently between 3 and 7
days, with a peak at day 5 since the onset of symptoms \cite{wolfelNature2020}.
The exact probability distribution for $\lambda$ remains an open issue
so our analysis assume a uniform distribution.

Finally, the SIR model rests on the random mixing hypothesis but
deviations are expected with stronger effects in population with
varying demographics or populations at risk.  In particular,
communicable respiratory diseases become major issues in correctional
facilities, given the lack of adequate environmental and sanitary
conditions. The combination of higher transmission rate of pathogens and
reduced removal rate with reduced healthcare  can increase the
crude mortality rate for incarcerated individuals. By a similar
argument, outbreaks in nursing homes may affect estimates of
epidemiological parameters because the disease becomes
disproportionately more lethal for older patients.
In this study, both effects are neglected in hopes to understand the
disease spreading of the average population with the minimal number of
parameters possible.

\section*{Acknowledgments}
The group belongs to the CNRS consortium ``Approches quantitatives du vivant''.

\section*{Author contributions statement}
GN wrote the paper and carried out the numerical analysis. CD, BG, and
MB designed the research and edited the paper.  All authors reviewed
the manuscript.   

\section*{Data availability}
The datasets analysed and numerical code generated during the current
study are available in the Zenodo repository, \url{https://doi.org/10.5281/zenodo.3931666}

\section*{Additional information} 
\textbf{Competing interests} The authors declare no conflict of interest.


\end{document}